\documentstyle[12pt]{article}
\advance\textheight by \topskip
\begin{document}
\begin{flushright}
SU-ITP-96-21\\
hep-th/9605112\\
\today\\
\end{flushright}
\vspace{1cm}
\begin{center}
\baselineskip=16pt

{\Large\bf    BRANELESS BLACK HOLES}  \\

\vskip 1.5 cm

{\bf Edi Halyo${}^{a}$\footnote{E-mail address:
 halyo@dormouse.stanford.edu},~
 Arvind Rajaraman${}^{b}$\footnote{E-mail address:
arvindra@dormouse.stanford.edu}~and Leonard Susskind${}^{a}$\footnote{E-mail
address:
 susskind@dormouse.stanford.edu}~}
\vskip 1 cm
${}^{a}$Department of Physics, Stanford University, Stanford CA 94305,
USA\\
\vskip 0.4 cm
${}^{b}$Stanford Linear Accelerator Center,\\
    Stanford University, Stanford, California 94309 USA
\\
\vskip 1.5cm
\end{center}
\begin{center}
{\bf Abstract}
\end{center}
It is known that the naive version of D-brane theory is inadequate to explain
the black hole
entropy in the limit in which the Schwarzschild radius becomes larger than all
compactification radii.
We present evidence that a more consistent description can be given in terms
of strings with rescaled tensions. We show that the rescaling can be
interpreted as
a redshift of the tension of a fundamental string in the gravitational field of
the black hole.
An interesting connection is found between the string level number and the
Rindler energy. Using this connection, we reproduce the entropies of
Schwarzschild black holes in
arbitrary dimensions in terms of the entropy of a single string at the Hagedorn
temperature.

\newpage

\section{Introduction}
Recently, a picture has been advocated\cite{SV}\cite{CM} of black hole entropy
as arising from
D-brane excitations.
This naive D--brane picture of black hole entropy is known to be inconsistent
when the
black hole becomes massive enough for its Schwarzschild radius to exceed any
microscopic scale such as the compactification radii. In \cite{MS} this was
called the limit of a
fat black hole.  A typical example analyzed by Callan and Maldacena\cite{CM}
and subsequently by Maldacena and Susskind\cite{MS} shows why this is so. In
this 5--dimensional example, five branes and one branes are wrapped on a five
torus and the system is given Kaluza--Klein momentum $N$ in one of the
directions. The D--brane picture says that the entropy is given in terms of a
partition function
\begin{equation}
Z=\prod_{n=1,\infty} \left({1+q^n \over 1-q^n }\right)^{4Q_1Q_5}= \sum
{d(N)q^N}
\end{equation}
for a gas of $Q_1Q_5$ species of massless quanta.
The integers $d(N)$ represent the degeneracy of the state with Kaluza-Klein
momentum number $N$.
For $N \to \infty$ keeping $Q_1Q_5$ fixed this gives an entropy
\begin{equation}
S=\log d(N) \to 2\pi \sqrt{NQ_1Q_5}
\end{equation}
This agrees with the classical black hole entropy. However, Maldacena and
Susskind
pointed out that the derivation is incorrect  in the case of fat black holes
because if $Q_1,Q_5$ and $N$
tend to infinity in fixed proportion, then one finds
\begin{equation}
\log d(N) \to N\log N
\end{equation}
which does not agree witth the black hole entropy. Furthermore the same formula
gives
\begin{equation}
\log d(N)=\log (Q_1Q_5)
\end{equation}
for fixed $N$ and large $Q_1Q_5$. Thus the naive D--brane model fails to agree
with
U--duality which requires symmetry among $Q_1,Q_5$ and $N$.

Maldacena and Susskind argued that a consistent theory could be formulated in
which
the $Q_1Q_5$ species are replaced by a single species and the level number $N$
is
replaced by $N^{\prime}=NQ_1Q_5$.  The entropy of the system is then carried by
a single long string with
a central charge six and a string tension $T \sim {1 \over {g
\alpha^{\prime}Q_5}}$. Evidently,
the picture which emerges from D--brane theory is that the black hole is a
single string
with a rescaled tension.  A similar picture has been advocated in the past by
one of us
\cite{S} and more recently by Tseytlin\cite{T}.

\section{A black hole in 5 dimensions}

We will analyze a 5-dimensional black hole made by wrapping $Q_5$ 5-D-branes on
$T^5$.
This was analyzed by Maldacena in \cite{M}. He showed that the near-extremal
entropy of
this configuration
is given by the entropy of closed strings which live on the 5--brane and
have zero winding and momentum. The central charge of these strings is given by
$c=6Q_5$ and their tension is $T={1 \over g\alpha'}$..  However, this system
suffers from the same problem described in
the introduction.The correct entropy is only obtained in the limit
$N=N_L=N_R>>c$
where $c$ is the central charge. However, for the near extremal case  $N$ is
small
compared to $c$. In  \cite{M}, it was conjectured that the correct
configuration is one in
which the string is fractionalized into strings with
\begin{equation}
c'=c/Q_5=6 \qquad \alpha'_{eff}=g \alpha' Q_5 \qquad N'=Q_5N
\end{equation}
 and the entropy formula is then correct. The rescaling of the string tension
gives
\begin{equation}\label{}
M=2\sqrt {N \over g \alpha' Q_5}=2\sqrt {N \over  \alpha'_{eff}}
\end{equation}
where $M$ is the deviation from extremality. The Hawking temperature becomes
\begin{equation}\label{}
T={1 \over \sqrt{g \alpha' Q_5}}={1 \over \sqrt{\alpha'_{eff}}}
\end{equation}
Note that the Hawking temperature appears to be the Hagedorn temperature
associated with the effective string scale.
We will show that the rescaling
of the string tension can be understood as the blueshift of the energy of the
string oscillations just
as the Hawking temperature can be understood as the redshifted Hagedorn
temperature \cite{S}.

 The metric in the transverse 5-dimensions is
 given in \cite{HS}.  We will use the notation of \cite{HMS} with
$\alpha=\sigma=0$. In this notation the
equation for the metric is
\begin{equation}
ds_5^2=-f^{-2/3}\left( 1-{r_0^2 \over r^2} \right) dt^2 +f^{1/3}\left[\left(
1-{r_0^2 \over r^2} \right)^{-1} dr^2 +r^2d\Omega_3^2 \right]
\end{equation}
where
\begin{equation}
f=\left( 1+{r_0^2sinh^2\gamma \over r^2} \right)
\end{equation}

The charge of the black hole  is
\begin{equation}
Q_5={r_0^2 \over 2g} sinh (2\gamma)\simeq {r_0^2 \over g} cosh^2 (\gamma)
\end{equation}
where the second relation is true in the near-extremal limit where
$r_0\rightarrow 0, \gamma\rightarrow \infty$.

We are interested in the near horizon limit where the (r,t) part of the metric
will be seen to be two-
dimensional Rindler spacetime. In this
limit
\begin{equation}
r\rightarrow r_0, f \rightarrow 1+sinh^2\gamma=cosh^2\gamma\equiv\lambda^3
\end{equation}
 To bring
the metric to the Rindler form, we rescale
\begin{equation}
r'=r\sqrt{\lambda} \qquad r'_0=r_0\sqrt{\lambda}
\end{equation}
Then
\begin{equation}
\lambda r'_0=r_0 cosh \gamma=\sqrt{gQ_5}
\end{equation}
Now expanding $r'=r'_0+y$, the near horizon metric takes the form
\begin{equation}
ds_5^2=-\lambda^{-2} {2y \over r_0'}dt^2+{r_0' \over 2y}dy^2+r'^2d\Omega_3^2
\end{equation}
 The proper distance $\rho$ from the horizon is then
\begin{equation}
\rho=\int{d\rho}=\int{\sqrt{r'_0 \over 2y}dy}=\sqrt{2r'_0}\sqrt{y}
\end{equation}
In terms of the proper distance, the coefficient of $dt^2$ becomes
\begin{equation}
g_{00}=-{\rho^2 \over \lambda^2 r'^2_0}
\end{equation}
Rescaling
\begin{equation}
\tau={t \over \lambda r'_0}
\end{equation}
the (r,t) part of the metric becomes of the Rindler form
\begin{equation}
ds_5^2=-\rho^2d\tau^2+d\rho^2
\end{equation}

The
 conjugate to $\tau$ is called the  Rindler energy and will scale inversely to
$\tau$. Hence we can write
\begin{equation}
E_R= \lambda r'_0\sqrt{\alpha'} M=M\sqrt{gQ_5\alpha'}
\end{equation}
Now we see a very interesting correspondence. Using eqn.(6) the Rindler energy
can
be simply written
\begin{equation}
E_R=2\sqrt N
\end{equation}
The Rindler energy is (apart from a factor of 2) just the square root of the
oscillator number !

We can now use the well-known entropy for a string with oscillator number N
\begin{equation}
S=2\pi\sqrt{{c \over 6}} 2\sqrt{N} = 2\pi\sqrt{{c \over 6}}E_R=2\pi\sqrt{{c
\over 6}}  M\sqrt{gQ_5\alpha'}
\end{equation}
For $c=6$, we find the temperature
\begin{equation}
T={1 \over 2\pi\sqrt{gQ_5\alpha'}}
\end{equation}
which is identical to the temperature found in \cite{M} .

Note that the redshift factor connecting $\tau$ and $t$ is exactly the ratio of
$\sqrt{\alpha'_{eff}}$
to $\alpha'$. Thus the picture is that of a fundamental string whose parameters
have been
rescaled due to redshifting.

\section{Schwarzschild Black Holes}

We now show that the same reasoning  holds for the Schwarzschild black hole
in any dimension $D$, $4\leq D \leq 10$. The Schwarzschild solution in $D$
space--time dimensions is given by (we use the notations of \cite{AP})
\begin{equation}
ds^2=-{r-\hat \mu \over r}dt^2+{r \over{r-\hat \mu}}dr^2+r^2d\Omega_2^2
\end{equation}
where
\begin{equation}
\rho=r^{D-3}\qquad\hat \mu ={16\pi G_NM \over{ (D-2)A_{D-2}}}
\end{equation}
Here $A_{D-2}={2\pi^{(D-1)/2} \over \Gamma\left({D-1 \over 2}\right)}$ is the
area of the unit sphere in $D-2$ dimensions. The black hole entropy is given by
\begin{equation}
S={M^{(D-2)/(D-3)} \over {4G_N}} A_{D-2}^{-1/(D-3)} \left(16\pi G_N \over {D-2}
\right)^{(D-2)/(D-3)}
\end{equation}
 Close to
the horizon, the proper distance to the horizon is given by
\begin{equation}
R={2 \over \sqrt {D-3}} \sqrt y \hat \mu^{1\over 2(D-3)}
\end{equation}
where $y$ is defined as $r=\hat \mu^{1/(D-3)}+y$ near the horizon. The
coefficient of the $dt^2$
term in the metric becomes
\begin{equation}
{r-\hat \mu \over r}={(D-3)^2 \over 4} R^2\left({16 \pi G_NM \over
(D-2)A_{D-2}}\right)^{-2/(D-3)}
\end{equation}
Rescaling $t$ to get the Rindler time $\tau$
\begin{equation}
\tau={(D-3) \over 2} \left({16 \pi G_NM \over (D-2)A_{D-2}}\right)^{-1/(D-3)} t
\end{equation}
The relation between $E_R$ and $M$ is more subtle in this case \cite{S}. The
Rindler energy may be
identified by requiring that it be conjugate to $\tau$, i.e.  $[E_R,\tau]=1$.
This can be written as
\begin{equation}
1= {(D-3) \over 2} \left({16 \pi G_NM \over
(D-2)A_{D-2}}\right)^{-1/(D-3)}[E_R,t]
\end{equation}
We now use the fact that $t$ is conjugate to $M$ to write
\begin{equation}
1= {(D-3) \over 2} \left({16 \pi G_NM \over (D-2)A_{D-2}}\right)^{-1/(D-3)}
{\partial E_R \over \partial M}
\end{equation}

The Rindler energy is then
\begin{equation}
E_R=M^{(D-2)/(D-3)} {2 \over D-2 }\left( {16 \pi G_NM \over
(D-2)A_{D-2}}\right)^{1/(D-3)}
\end{equation}
Now the relation derived in the previous section between $S$ and $E_R$
\begin{equation}
S=2\pi E_R \sqrt{c \over 6}
\end{equation}
gives the correct black hole entropy for $c=6$. Note that $c$ must be $6$ in
all dimensions.
We have no deep understanding of why this is so, but at least in the cases of
$D=4,5$,
this is supported by the analysis of Tseytlin\cite{T}.
In addition, the identification of the square root of the string oscillator
number $N$ with the Rindler energy $E_R$ works in all dimensions.

\section{String Length and Entropy}

We would like to make one more comment on the relation of entropy and string
theory,
which may be relevant to the question of black hole entropy. The point concerns
a connection between the entropy carried by a string and its
total integrated length.

Consider a free string thermally excited to average level $N$. Its entropy is $
2\pi\sqrt{{Nc \over 6}}$.

Let us consider the total transverse length of the string. It is given in
light-cone frame by
\begin{equation}
L=\langle \int_{0}^{2\pi} d\sigma  \sqrt{ \partial_\sigma X_i\partial_\sigma
X_i} \rangle
= 2\pi\langle \sqrt{ \partial_\sigma X_i\partial_\sigma X_i} \rangle
\end{equation}

Now it is easy to prove that for a linear system like a free string that
\begin{equation}
\langle \sqrt{ \partial_\sigma X_i\partial_\sigma X_i} \rangle\propto \langle
\partial_\sigma X_i\partial_\sigma X_i \rangle^{1/2}
\end{equation}

Furthermore, by the virial theorem,
\begin{equation}
\langle  \partial_\sigma X_i\partial_\sigma X_i \rangle=N\alpha'
\end{equation}
where we have dropped quantum fluctuations.

Thus the mean length of string is
\begin{equation}
L \sim N^{1/2}\sqrt{\alpha'}
\end{equation}

On the other hand, the entropy of the string is $N^{1/2}$. Thus we find that
the entropy
per unit length
of the string is a universal constant of order 1 bit per string length. If we
think of the string as being
subdived into segments of size $\sqrt{\alpha'}$, then the entropy is the number
of
segments. This suggests that the horizon is occupied by string segments with
a universal density of 1 segment per Planck area.

\section{Discussion}

As far as we can tell, the entropy of fat black holes seems to have little
to do with D-branes and more to do with string-like degrees of freedom.
The entropy is dominated by configurations containing one long string.
D-branes do seem to be important in certain limits, for example when the
Kaluza-Klein momentum N
is much larger than the other charges. This for example can be achieved when
one of the compactification radii is much larger than the others and the black
hole becomes a black string.

In the fat black hole limit,  the same string like degrees of freedom seem to
be
relevant for near extremal black holes as well as the opposite limit of
Schwarzschild black holes. This suggests that they may also be relevant
for other black holes.

\section{Acknowledgements}

The work of A.R is supported in part by the Department of Energy under contract
number DE-AC03-76SF00515. L.S. was supported in part by NSF grant  PHY 9219345.

\end{document}